\documentclass[aps,prd,twocolumn,showpacs,preprintnumbers,amsmath,amssymb,superscriptaddress,nofootinbib]{revtex4-2}

\usepackage{graphicx}
\usepackage{amsmath}
\usepackage{amssymb}
\usepackage{dcolumn}
\usepackage{bm}
\usepackage{xcolor}
\usepackage{float}
\usepackage{comment}
\usepackage[normalem]{ulem}
\usepackage[colorlinks=true,linkcolor=blue!45!black,citecolor=blue!45!black,
            urlcolor=blue!45!black]{hyperref}
\newcommand{\Mpl}{M_{\rm pl}}

\newcommand{\beq}{\begin{equation}}
\newcommand{\eeq}{\end{equation}}
\newcommand{\rhophi}{\rho_\phi}
\newcommand{\Ueff}{U_{\rm eff}}
\newcommand{\Ukin}{U_{\rm kin}}
\newcommand{\Neff}{\mathcal{N}}
\begin{document}

\title{Kinetic backreaction cannot suppress axion quantum pressure}

\author{Kaleb Anderson}
\email{kaleb\_anderson@brown.edu}
\affiliation{Department of Physics and Brown Center for Theoretical Physics \& Innovation, Brown University, Providence, RI 02912-1843, USA}

\author{Savvas M.~Koushiappas}
\email{savvas\_koushiappas@brown.edu}
\affiliation{Department of Physics and Brown Center for Theoretical Physics \& Innovation, Brown University, Providence, RI 02912-1843, USA}

\date{\today}

\begin{abstract}

Ultralight axions from string compactifications carry non-canonical kinetic 
terms whose normalization is set by a coupling function that depends on an 
accompanying modulus field. As dark matter, such an axion forms a cold 
condensate whose quantum pressure resists gravitational collapse below a Jeans 
scale fixed by that coupling. One might hope that backreaction from the 
condensate's oscillations could shrink this Jeans length and thereby enhance 
structure growth. We show that it cannot. For any smooth, positive coupling 
function, the oscillation-averaged backreaction always drives the coupling to 
larger values, strengthening the quantum pressure and growing the Jeans length 
with time. We establish this no-go result analytically through an adiabatic-
invariant analysis and confirm it by direct numerical integration. Kinetically 
coupled dark matter therefore cannot self-generate the conditions for enhanced 
gravitational collapse or the seeding of structure on sub-Jeans scales.

\end{abstract}

\pacs{98.80.Cq, 95.35.+d, 04.70.Dy, 04.30.Db}

\maketitle

\section{Introduction}
\label{sec:intro}

Ultralight bosons with masses $m\sim10^{-22}\text{--}10^{-20}\, \mathrm{eV}$ are well-motivated alternatives to cold dark matter on small scales~\cite{Preskill:1982cy, Hu2000,Hui2017,Marsh2016,Abbott:1982af,Dine:1982ah}. {In the presence of appropriate self-interaction,} such ``fuzzy'' dark matter behaves as a cold condensate on large scales but resists gravitational clustering below its de Broglie wavelength \cite{Guth:2014hsa}. In the fluid (Madelung) description, this appears as a quantum pressure that competes with self-gravity and defines a Jeans wavenumber
\begin{equation}
    {\left( \frac{k_J}{a} \right)^2 = \sqrt{6H^2 m^2 \Omega_\phi}}
    \label{eq:kJ_standard}
\end{equation}
above which perturbations oscillate rather than grow \cite{Hui2017, Guth:2014hsa}. This scale is the signature of ultralight dark matter, suppressing the matter power spectrum \cite{Hu2000, Marsh2016, Schive:2014hza, Hlozek:2014lca, Marsh:2015wka}, flattening galactic cores \cite{Schive:2014dra, Schive:2014hza,Goldstein:2022pxu}, and setting a minimum halo mass \cite{Marsh:2013ywa, Schive:2015kza}.

Light scalars of this type arise naturally in string compactifications, where axions appear from the dimensional reduction of higher-form gauge fields~\cite{Svrcek2006,Arvanitaki2010} and generically lack canonical kinetic terms. Their normalization is fixed by the internal geometry and hence depends on the moduli. The relevant part of the four-dimensional action is
\begin{equation}
  \mathcal{L}\supset
  -\dfrac12(\partial\chi)^2-\dfrac12 f(\chi)(\partial\phi)^2,
  \label{eq:Lkin}
\end{equation}
with $\phi$ the axion, $\chi$ a canonically normalized modulus, and $f(\chi)>0$ a kinetic-coupling function fixed by the K\"ahler potential (typically an exponential $f(\chi)=e^{\lambda\chi}$ with $\lambda=\mathcal{O}(1)$). This kinetic mixing is present whenever the modulus is not stabilized above the relevant scale, and has been used to build string-motivated dark-sector models---for example, by transferring energy from an early-dark-energy axion into a stiff modulus to address the Hubble and $S_8$ tensions~\cite{Alexander2019,KMIX2022}.

The coupling in Eq.~\eqref{eq:Lkin} has an interesting consequence for the condensate's quantum pressure. Because the quantum-pressure term originates from the axion kinetic structure, rescaling that structure by $f(\chi)$ rescales the pressure, and the Jeans wavenumber of
Eq.~\eqref{eq:kJ_standard} becomes~\cite{Toomey2025}
\begin{equation}
  \left(\frac{\tilde k_J}{a}\right)^2
  =\sqrt{{\cal{F}}^2+\frac{6H^2m^2\Omega_\phi}{f}}-{\cal{F}},
  \label{eq:kJ_mod}
\end{equation}
where ${\cal{F}} \equiv m^2(f-1)/2f$. Eq.~\eqref{eq:kJ_mod} then reduces to Eq.~\eqref{eq:kJ_standard} when $f\to1$. The main point is the monotonic dependence on $f$. In the case where $f>1$, the dynamical effect of the quantum pressure is enhanced relative to gravity, allowing structure to form on larger scales, as was explored in \cite{Toomey2025}. Conversely, Eq.~\eqref{eq:kJ_mod}  suggests that for $f<1$, the effect is reversed so that the Jeans length decreases.

The Jeans scale thus inherits the dynamics of the modulus, and the axion's energy density sources the modulus through the kinetic coupling, i.e., the modulus responds to the condensate. If that "backreaction" could drive $f$ below unity during the radiation era, perturbations previously supported by quantum pressure would become unstable and collapse, potentially accelerating early structure formation and, speculatively, aiding in the seeding of supermassive black holes. If, instead, the backreaction drives $f$ above unity, small-scale structure is suppressed further. Which way the backreaction points is therefore a basic dynamical question for any kinetically coupled condensate, and bears directly on the structure-formation constraints through which such models are tested~\cite{Toomey2025}.

In this paper, we show that the backreaction \emph{always} drives $f$ upward and can never dynamically suppress the quantum pressure. The conclusion does not depend on the sign of $\lambda$, the modulus potential, or initial conditions; it is a fundamental property of the coupling~\eqref{eq:Lkin} that holds for \emph{every} smooth $f(\chi)>0$. 

For the exponential coupling $f=e^{\lambda\chi}$, Alexander, Bernardo, and Toomey~\cite{KMIX2022} showed that the oscillation-averaged modulus rolls in an effective potential, and explored the resulting axion-modulus energy transfer to dilute an early-dark-energy axion and address the Hubble and $S_8$ tensions (building on the axio-dilaton dynamics of Ref.~\cite{Alexander2019}). We build on that result in two ways. First, the conclusion holds for \emph{arbitrary} smooth $f(\chi)>0$, not only the monotonic exponential: the oscillation-averaged drift obeys $\dot f\propto[f'(\chi)]^2\ge0$, a form that is independent of the shape of $f$ and shows that non-monotonic couplings drive $\chi$ to the maxima of $f$. Second, and more importantly, we connect these dynamics to the condensate quantum pressure: the modified Jeans relation derived (for static $\chi$) in Ref.~\cite{Toomey2025} gives\footnote{The scaling $\tilde{k}_J \propto f^{-1/2}$ refers to the $m \gg H$ regime, in which ${\cal{F}}^2 \gg 6 H^2 m^2 \Omega_\phi / f$ and Eq. \eqref{eq:kJ_mod} reduces to $(\tilde{k}_J / a)^2 \approx 6 H^2 \Omega_\phi / f$; this is the regime relevant to structure formation.} $\tilde k_J\propto f^{-1/2}$, so the monotonic increase of $f$ fixes the sign of $d\tilde k_J/dt$ and forbids the self-suppression of quantum pressure that the rescaled Jeans relation might otherwise seem to permit.

The paper is structured as follows. In Sec.~\ref{sec:setup} we write the model, averaged over the rapid axion oscillations using the virial theorem and the adiabatic invariant, and reduce the modulus dynamics to a motion in an effective potential. In Sec.~\ref{sec:no-go}, we establish the central result---the axion-generated part of that potential decreases monotonically in $f$, with the no-go result encoded in the perfect square $\dot f\propto[f']^2$, and show that no choice of modulus potential can rescue the mechanism. In Sec.~\ref{sec:numerics}, we confirm the result by integrating the full unaveraged equations, and in Sec.~\ref{sec:discussion}, we discuss its scope, its relation to integrable limits of kinetically mixed dark sectors, and the assumptions a working mechanism would have to break. Throughout, we use natural units $\hbar=c=1$, signature $(-,+,+,+)$, and reduced Planck units $8\pi G=\Mpl^{-2}$ where convenient; a dot denotes a derivative with respect to cosmic time $t$, $H\equiv\dot a/a$, and a prime on a function of a single field denotes a derivative with respect to that field.

\section{Model, averaging, and the effective potential}
\label{sec:setup}

We consider a single ultralight axion $\phi$ kinetically coupled to a single canonically normalized modulus $\chi$ through the two-field action studied in Refs.~\cite{Alexander2019,KMIX2022,Toomey2025},
\begin{equation}
  S=\int d^4x\,\sqrt{-g}\left[
    -\dfrac12(\partial\chi)^2
    -\dfrac12 f(\chi)(\partial\phi)^2
    -V(\chi)-U(\phi)\right],
  \label{eq:action}
\end{equation}
with $f(\chi)>0$ a smooth but otherwise arbitrary kinetic-coupling function, $U(\phi)=m^2 f_a^2[1-\cos(\phi/f_a)]$ the axion potential with $U''(0)=m^2$, and $V(\chi)$ an arbitrary modulus potential, discussed in Sec.~\ref{sec:no-go}. Unlike previous treatments, we make no assumption on the sign, monotonicity, or functional form of $f$.

For homogeneous fields in a spatially flat Friedmann--Lema\^itre--Robertson--Walker background, the equations of motion that follow from Eq.~\eqref{eq:action} are \cite[Eq.~2.3]{KMIX2022}
\begin{align}
  \ddot\chi+3H\dot\chi-\dfrac12 f'(\chi)\,\dot\phi^2+V'(\chi)&=0,
  \label{eq:chi}\\[2pt]
  \ddot\phi+\left(3H+\frac{f'(\chi)}{f(\chi)}\dot\chi\right)\dot\phi
    +\frac{1}{f(\chi)}U'(\phi)&=0.
  \label{eq:phi}
\end{align}
The modulus equation, Eq.~\eqref{eq:chi}, contains a \emph{source} term $-\tfrac12 f'(\chi)\dot\phi^2$ proportional to the axion kinetic energy so that a moving axion exerts a force on $\chi$, with the sign fixed by the action. Meanwhile, the axion equation, Eq.~\eqref{eq:phi}, contains a \emph{friction-like} term $(f'/f)\dot\chi\,\dot\phi$ through which energy flows between the fields. The question is which way the source drives $\chi$: toward $f>1$ and increased quantum pressure, or toward $f<1$ and suppressed quantum pressure.

In the coupled form~\eqref{eq:chi}--\eqref{eq:phi}, the rapidly
oscillating source obscures the net direction of the force on $\chi$. Our goal is therefore to recast the modulus dynamics as motion with friction in a fixed effective potential $\Ueff(\chi)$: a field rolling with friction settles at the potential's minima, so the entire problem reduces to identifying the extrema of a single function, for \emph{any} $f$, rather than solving two coupled nonlinear equations. We do this in three steps. First, we average the source over one axion oscillation, using the virial theorem to express $\langle\dot\phi^2\rangle$ through the axion energy density $\rhophi$. Second, we fix the dependence of $\rhophi$ on the modulus and the scale factor from the adiabatic invariance of the oscillation. Finally, we show that the averaged source is a total $\chi$-derivative that can be absorbed into the potential.

The averaging is valid when the axion oscillates rapidly compared to the background evolution---the same regime in which the condensate, and hence its quantum pressure, is defined. Writing the effective axion mass as $m_{\rm eff}\equiv m/\sqrt{f}$ (the factor of $f^{-1/2}$ follows from canonically normalizing the axion at fixed $\chi$, $\phi_c=\sqrt{f}\,\phi$, so that $\tfrac12 m^2\phi^2=\tfrac12(m^2/f)\phi_c^2$), the relevant conditions are
\begin{equation}
  m_{\rm eff}\gg H,
  \qquad
  m_{\rm eff}\gg\left|\frac{\dot f}{f}\right|.
  \label{eq:adiab}
\end{equation}

The first condition states that the axion completes many oscillations per Hubble time, the second that the kinetic coupling varies slowly over an oscillation. Both hold for the ultralight masses and slowly rolling moduli of interest, so we may average over the fast oscillation while treating $\chi$, $H$, and $f$ as constant within a cycle. Since the condition in Eq.~\eqref{eq:adiab} depends on the effective mass, which is a function of the kinetic coupling, there is not a {well-defined} energy regime where this condition holds\footnote{Quantitatively, $m_{\rm eff} \gg H$ with $H \sim T^2/m_{\rm pl}$ requires $T \le \sqrt{m_{\rm eff} m_{\rm pl}}$, i.e., temperatures below $\sim$ a few hundred eV to a few keV across $m \sim 10^{-22} - 10^{-20}$ eV; this condensate description applies throughout the relevant structure formation era.}. 
%

Within one oscillation, the axion energy density is $\rhophi=\tfrac12 f\dot\phi^2+\tfrac12 m^2\phi^2$, and the virial theorem for a harmonic oscillator\footnote{Anharmonic corrections to the cosine potential are negligible for the small-amplitude oscillations relevant here. The misalignment amplitude redshifts as $\phi \propto a^{-3/2} f^{-1/4}$, so $\phi / f_a \ll 1$ throughout, and $\langle {\rm KE} \rangle = \langle {\rm PE} \rangle$ holds to leading order.} gives $\langle\tfrac12 f\dot\phi^2\rangle=\tfrac12\rhophi$. Since the source term in Eq.~\eqref{eq:chi} is $\tfrac12 f'\dot\phi^2=(f'/f)\,\tfrac12 f\dot\phi^2$, averaging yields
\begin{equation}
  \left\langle\dfrac12 f'(\chi)\dot\phi^2\right\rangle
  =\frac{f'(\chi)}{2f(\chi)}\,\rhophi.
  \label{eq:src}
\end{equation}
That is, the rapidly oscillating $\dot\phi^2$ is replaced, on average, by a smooth function of the slowly evolving energy density.
Inserting the averaged source, Eq.~\eqref{eq:src}, into
Eq.~\eqref{eq:chi} gives the oscillation-averaged dynamics
\begin{equation}
  \ddot\chi+3H\dot\chi-\frac{f'(\chi)}{2f(\chi)}\,\rhophi(\chi,a)
  +V'(\chi)=0.
  \label{eq:chiavg}
\end{equation}
Note that the averaged backreaction term is itself the gradient of a function of $\chi$.

The dependence of $\rhophi$ on $\chi$ and $a$ is fixed by the adiabatic invariant of the simple harmonic oscillator. For the canonically normalized axion of mass $m_{\rm eff}$, conservation of the oscillator action $\oint p\,dq$ is conservation of comoving particle number \cite{Turner:1983he},
\begin{equation}
  \Neff\equiv\frac{\rhophi\,a^3}{m_{\rm eff}}=\text{const}.
  \label{eq:adiabatic_invariant}
\end{equation}
The number of axion quanta in a comoving volume is unchanged by the adiabatic evolution of expansion or coupling, and only the energy per quantum $m_{\rm eff}=m/\sqrt f$ responds. Hence,
\begin{equation}
  \rhophi(\chi,a)
  =\frac{\Neff\,m_{\rm eff}}{a^3}
  \equiv\frac{C(a)}{\sqrt{f(\chi)}}.
  \label{eq:rho}
\end{equation}
The prefactor, $C(a)=\Neff\,m/a^3$, is positive and redshifts as $a^{-3}$ so the entire $\chi$-dependence of the averaged axion energy density resides in the factor $f(\chi)^{-1/2}$. Direct differentiation gives
\begin{equation}
  \frac{\partial}{\partial\chi}\rhophi(\chi,a)
  =-\frac{f'(\chi)}{2f(\chi)}\,\rhophi.
  \label{eq:identity}
\end{equation}
This implies that the averaged axion backreaction on the modulus is exactly the field-space gradient of the averaged axion energy density (this holds whenever the oscillation averaging and the adiabatic invariant Eq.~\eqref{eq:adiabatic_invariant} apply). Therefore, the backreaction is conservative, not dissipative, and the modulus feels it as an ordinary force derived from a potential.

The averaged modulus equation~\eqref{eq:chiavg} therefore describes a single degree of freedom rolling with Hubble friction in an effective potential\footnote{This is an energy-density-sourced effective potential analogous to what appears in chameleon~\cite{Khoury:2003rn} and coupled-quintessence~\cite{Amendola:1999er} dynamics.},
\begin{equation}
  \ddot\chi+3H\dot\chi+\Ueff'(\chi,a)=0,
  \label{eq:eff}
\end{equation}
with
\begin{equation}
  \Ueff(\chi,a)=V(\chi)+\Ukin(\chi,a),
  \qquad
  \Ukin\equiv\frac{C(a)}{\sqrt{f(\chi)}}.
  \label{eq:Ueff}
\end{equation}
The axion enters only through $\Ukin\propto f(\chi)^{-1/2}$, its averaged energy density viewed as a potential for $\chi$. Equation~\eqref{eq:eff} is the form we sought: the oscillating two-field backreaction problem has become a one-dimensional particle rolling in a fixed (slowly redshifting) potential well.

\section{The no-go result}
\label{sec:no-go}
Everything now follows from the shape of the kinetic effective potential $\Ukin(\chi)\propto f(\chi)^{-1/2}$. Differentiating with respect to the modulus, the axion contributes to Eq.~\eqref{eq:eff} the force $-\Ukin'(\chi,a)$ with
\begin{equation}
  \Ukin'(\chi,a)
  =-{\cal{C}}\,f(\chi)^{-3/2}f'(\chi),
  \label{eq:Up}
\end{equation}
where ${\cal{C}} = C(a)/2$. Three consequences follow. First, the direction of the force. Because $C(a)>0$ and $f>0$, the prefactor ${\cal{C}}f^{-3/2}$ is strictly positive, so the backreaction force on $\chi$ carries the sign of $f'(\chi)$ everywhere. Where $f'>0$, the force pushes $\chi$ up, increasing $f$; where $f'<0$ it pushes $\chi$ down, again increasing $f$. In both cases, the modulus is driven toward \emph{increasing} $f$.
Second is where in field space the modulus comes to rest. The stationary points of $\Ukin$ are the solutions of $f'(\chi)=0$ (the stationary points of $f$ itself). Differentiating Eq.~\eqref{eq:Up} once more and evaluating where $f'=0$ so that the term proportional to $(f')^2$ drops out,
\begin{equation}
  \Ukin''(\chi,a)\big|_{f'=0}
  =-{\cal{C}}f^{-3/2}f''(\chi).
  \label{eq:Upp}
\end{equation}
The sign of $\Ukin''$ is opposite to that of $f''$: a stationary point is a \emph{minimum} of $\Ukin$ precisely when it is a \emph{maximum} of $f$, while the minima of $f$ are maxima of $\Ukin$ and dynamically unstable. Therefore, the modulus relaxes toward the maxima of $f$, but never at the minima.

\begin{figure*}[t]
  \centering
  \includegraphics[width=2\columnwidth]{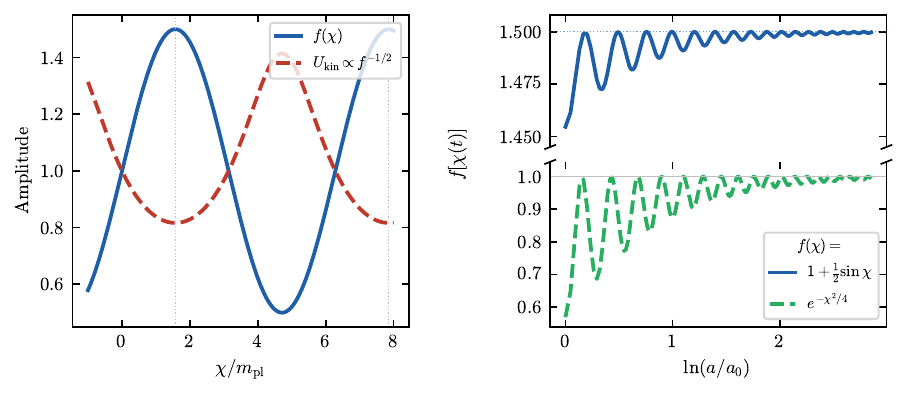}
  \caption{Numerical confirmation of the no-go result of
    Sec.~\ref{sec:no-go}.
    {\it{Left}}: For the non-monotonic coupling
    $f(\chi)=1+\tfrac12\sin\chi$, the kinetic effective potential $\Ukin\propto f^{-1/2}$ (dashed) is minimized exactly at the maxima of $f$ (solid); the oscillation-averaged dynamics drive $\chi$ toward those minima of $\Ukin$, i.e.\ the maxima of $f$. {\it{Right}}: Full unaveraged integration of the background system~\eqref{eq:chi}--\eqref{eq:phi} in a radiation background, for $f=1+\tfrac12\sin\chi$ and $f=e^{-\chi^2/4}$. $a_0$ corresponds to an arbitrary initial scale factor. In both cases, $f(\chi(t))$ rises in the mean toward its maximum, confirming that the backreaction increases $f$ independent of the functional form of the coupling.}
  \label{fig:nogo}
\end{figure*}

Third, and more importantly, is the rate at which $f$ changes along the trajectory. In the friction-dominated (slow-roll) regime we drop $\ddot\chi$ in Eq.~\eqref{eq:eff} and take $V$ negligible, so that $3H\dot\chi=-\Ukin'={\cal{C}}f^{-3/2}f'$. Then,
\begin{align}
  \dot f
  &=f'(\chi)\,\dot\chi
  =f'(\chi)\cdot\frac{1}{3H}\!\left({\cal{C}}f^{-3/2}f'\right)
  \nonumber\\
  &=\frac{C(a)}{6H}\,f^{-3/2}\,[f'(\chi)]^2\;\ge\;0,
  \label{eq:fdot}
\end{align}
with equality only at the stationary points of $f$. The oscillation-averaged backreaction \emph{never lowers $f$ for any smooth $f(\chi)>0$}. The rate is controlled by the perfect square $[f'(\chi)]^2$ so that the sign of the rate does not depend on a particular coupling. In particular, the force on $\chi$ is proportional to $f'(\chi)$ [Eq.~\eqref{eq:Up}], and the response of $f$ to a change in $\chi$ is \emph{also} proportional to $f'(\chi)$ by the chain rule, so their product inevitably carries two factors of $f'$ and cannot change sign. No field redefinition and no choice of $f$ can evade this because force and response transform the same way.

This is the main result of the paper. The modified Jeans wavenumber scales as $\tilde k_J\propto f^{-1/2}$, so a non-decreasing $f$ means a non-increasing $\tilde k_J$: the condensate quantum pressure is driven \emph{up}, not down, by the axion's own backreaction. Starting from $f\simeq1$, the system evolves towards $f\ge1$, enlarging the Jeans length. The dynamical reduction to $f<1$ simply cannot occur through an energy exchange between the axion and the moduli field.

One might still hope that a suitably chosen modulus potential $V(\chi)$ could carry $\chi$ into the $f<1$ region in spite of the backreaction, since the full effective potential is $\Ueff=V+\Ukin$. No such rescue is viable. If $V$ is a runaway potential---the generic situation for a logarithmically stabilized string modulus \cite{Burgess:2021obw, Bernardo:2022ztc}, $V\propto e^{-\alpha\chi}$ with $\alpha>0$---its gradient pushes $\chi$ monotonically in one direction without bound. For a coupling with $f'<0$ in that direction, this can carry $\chi$ into a region with $f<1$, but with no minimum to halt it and no force to return $\chi$ to $f\simeq1$. In this scenario, once $f \ll 1$, the Jeans wavelength shrinks to the point where it is no longer sensible to talk about a soliton core since only the highest modes are stable against collapse. Here, the condensate remains; however, the effective mass continues to grow until macroscopic behavior that made ultralight dark matter interesting is lost.

Moreover, this behavior is a property of the bare modulus potential and would occur with or without the axion; it is not the mechanism in which the axion oscillations drive the suppression (which is what the Alexander paper did). The backreaction only hinders the approach to $f<1$. By Eq.~\eqref{eq:Up}, wherever reaching $f<1$ requires moving against $f'$, the term $\Ukin$ contributes a restoring force of magnitude ${\cal{C}}f^{-3/2}|f'|$ that grows without bound as $f\to0$. The backreaction thus acts as a barrier that steepens without limit toward exactly the region the mechanism must reach, and any $V$ that gets there does so in spite of the axion, not because of it. 

If instead $V$ is stabilized, with a genuine minimum at some $\chi_\star$ (as in the Racetrack \cite{Kallosh:2004yh, Blanco-Pillado:2005arr, Kallosh:2014oja} or Large Volume \cite{Balasubramanian:2005zx, Cicoli:2008va} constructions), the late-time modulus settles there, and the coupling is fixed at $f(\chi_\star)$. A viable late-time condensate requires $f(\chi_\star)\simeq1$; but then the backreaction pushes $\chi$ toward the nearest \emph{maximum} of $f$ while the potential restores it to $\chi_\star$. Consequently, the trajectory never makes the large excursion into $f<1$ that a dynamical suppression would require. Tuning $\chi_\star$ so that $f(\chi_\star)<1$ permanently is a static property of the chosen vacuum, and not a dynamical suppression sourced by the axion. 

\section{Numerical confirmation}
\label{sec:numerics}

The argument of Sec.~\ref{sec:no-go} relies on oscillation averaging and the adiabatic invariant. To verify that approach, we integrate the \emph{full unaveraged} background system~\eqref{eq:chi}--\eqref{eq:phi} numerically, retaining the rapid axion oscillations explicitly. We work in a radiation-dominated background, $H=1/(2t)$, with the axion mass large compared to $H$ throughout the integration. We then take $V=0$ to isolate the kinetic backreaction. Finally, we consider three qualitatively distinct couplings: a non-monotonic oscillatory coupling $f(\chi)=1+\tfrac12\sin\chi$; a bell-shaped coupling with an interior maximum, $f(\chi)=e^{-\chi^2/4}$; and the exponential couplings $f(\chi)=e^{\pm\lambda\chi}$ of the string-motivated case.

Figure~\ref{fig:nogo} summarizes the result. In every case $f(\chi(t))$ rises in the mean and asymptotes toward a local maximum of $f$, never decreasing persistently, in agreement with Eqs.~\eqref{eq:Up}--\eqref{eq:fdot}. Notably, a trajectory initialized where $f$ is locally \emph{decreasing}---where one might naively expect the field to slide toward smaller $f$---still reverses and climbs to the adjacent maximum. The exponential cases $f=e^{\pm\lambda\chi}$ (not plotted) drive $\chi\to\pm\infty$ monotonically with $f\to\infty$, consistent with the absence of any interior maximum. 

\section{Discussion and conclusions}
\label{sec:discussion}

We have shown that an ultralight axion kinetically coupled to a string modulus through any smooth, positive $f(\chi)$ cannot dynamically suppress its own condensate quantum pressure. The axion enters the modulus dynamics only through its energy density, $\rhophi\propto f(\chi)^{-1/2}$: a larger $f$ means a heavier effective axion but, at fixed comoving number, a \emph{lower} energy density. The modulus rolls so as to lower the energy it sees, hence toward larger $f$. There is no way for the field to use its own positive kinetic energy to shrink the coefficient multiplying that energy.

The effective-potential structure itself is not new: a scalar acquiring a density-sourced effective potential whose extrema tracks the ambient energy density is the mechanism underlying chameleon~\cite{Khoury:2003rn}, symmetron \cite{Hinterbichler:2010es} and coupled-quintessence~\cite{Amendola:1999er} models. In the flat-space, potential-free, integrable limit, the conserved momentum $p=f(\chi)\dot\phi$ gives $\tfrac12\dot\chi^2+\tfrac12 p^2 f^{-1}=E$, so $\chi$ moves in a potential $\propto p^2 f^{-1}$---the exact-momentum analogue of our averaged $\Ukin\propto f^{-1/2}$, with the same monotonic decrease in $f$. This effective-potential description, and the conclusion that kinetic mixing transfers energy from the axion to the modulus, were established for the exponential coupling in Refs.~\cite{Alexander2019,KMIX2022}. What we add is the generalization to arbitrary smooth $f(\chi)>0$, with the attractors identified as the maxima of $f$, and the consequence for the condensate quantum pressure: since $\tilde k_J\propto f^{-1/2}$~\cite{Toomey2025}, the monotonic growth of $f$ forbids any dynamical self-suppression of the Jeans scale. In short, the very property that makes kinetic mixing a reliable energy-transfer mechanism is what makes it useless as a quantum-pressure switch.

The direction of the kinetic backreaction is therefore fixed: it drives the modulus so as to strengthen, never weaken, the condensate quantum pressure. This addresses a basic dynamical question for kinetically coupled axion--moduli dark matter. The same energy-transfer property that makes kinetically mixed dark sectors attractive for other cosmological purposes is precisely what prevents them from switching off the quantum pressure of their own dark matter.

\section*{Acknowledgments}
\noindent
We acknowledge useful conversations with Stephon Alexander, Alexis Ortega, and Michael Toomey. We acknowledge the use of the Claude AI Assistant (Anthropic) for coding assistance and in checking the calculations in this manuscript. The authors take sole responsibility for all content. This work was supported by National Science Foundation (NSF) Grant No. PHY-2412666.

\end{document}